\documentclass{moriond}
\usepackage{color}
\usepackage{amsmath}
\usepackage{calrsfs}
\usepackage{graphicx,wrapfig,lipsum}

\def\be{\begin{equation}}
\def\ee{\end{equation}}
\def\bea{\begin{eqnarray}}
\def\eea{\end{eqnarray}}

\definecolor{darkpastelgreen}{rgb}{0.01, 0.75, 0.24}

\newcommand{\red}[1]{\textcolor{red}{#1}}
\newcommand{\blue}[1]{\textcolor{blue}{#1}}

\newcommand{\green}[1]{\textcolor{darkpastelgreen}{#1}}

\newcommand{\Photo}{\includegraphics[height=35mm]{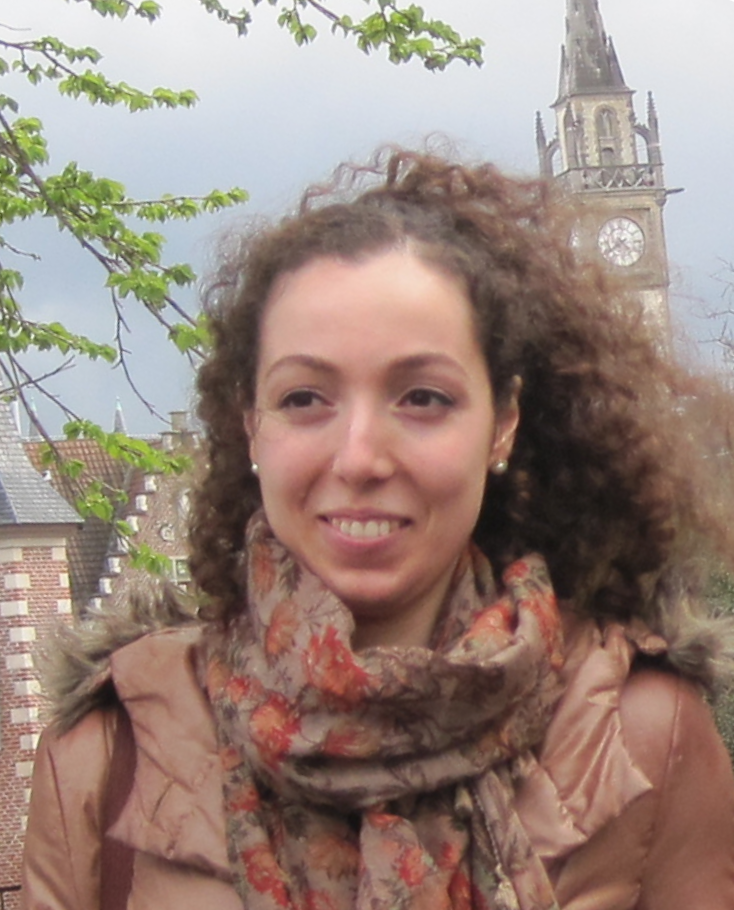}}

\begin{document}
\vspace*{4cm}
\title{SEARCH FOR SHARP NEUTRINO FEATURES FROM DARK MATTER DECAY \cite{Aisati:2015vma}}

\author{ C. EL AISATI }

\address{Service de Physique Th\'eorique - Universit\'e Libre de Bruxelles, \\
Boulevard du Triomphe, CP225, 1050 Brussels, Belgium}

\maketitle\abstracts{The discovery of a neutrino line or, more broadly, a sharp feature in neutrino data could provide a striking hint for the existence of the dark matter particle. We review here a search for sharp spectral features using neutrino data from IceCube. No significant hint for a signal from decaying dark matter was found in the analysed dataset. Yet, we show a strong improvement of the limits, which makes it worth discussing how they compare to the limits obtained from gamma-ray data. 
}

\section{Introduction}
Because of gravitational evidences at \textit{all} scales, the existence of dark matter (DM) in our Universe is considered well established. Yet, the nature of the second main ingredient of the cosmic cocktail remains a mystery. According to some theoretical models, DM particles could annihilate or decay, and produce sharp features in the gamma-ray and/or neutrino spectra (e.g. lines in case of $\gamma \gamma$ and $\nu \bar{\nu}$ finales states, to name but a few). In the context of indirect searches these features are dubbed as ``smoking guns'', for they cannot be mimicked by conventional astrophysical backgrounds at multi-GeV energies, and hence constitute the most striking signatures of DM to look for.

In the following, the focus is put on neutrino features. Since the discovery of an extra-terrestrial flux in 2013, neutrinos have gained in popularity but our motivation to study their possible features also lies in the fact that an energy resolution comparable to the one of current gamma-ray telescopes in the GeV to TeV energy range ($10-15\%$) is achievable by IceCube (e.g. for a fully contained event selection) \cite{Aartsen:2013vja}. This, together with the fact that many years of data are becoming available, could potentially provide sensitivities competing with analogous searches in the gamma-ray sky.

We propose to review the broad lines of our analysis in Sec.~\ref{sec:statanalysis}. Results are presented in Sec.~\ref{sec:results}, and are followed by a summary. 

\section{Statistical analysis in a nutshell}
\label{sec:statanalysis}
\begin{figure}[h!]
\centering
\includegraphics[scale=.4]{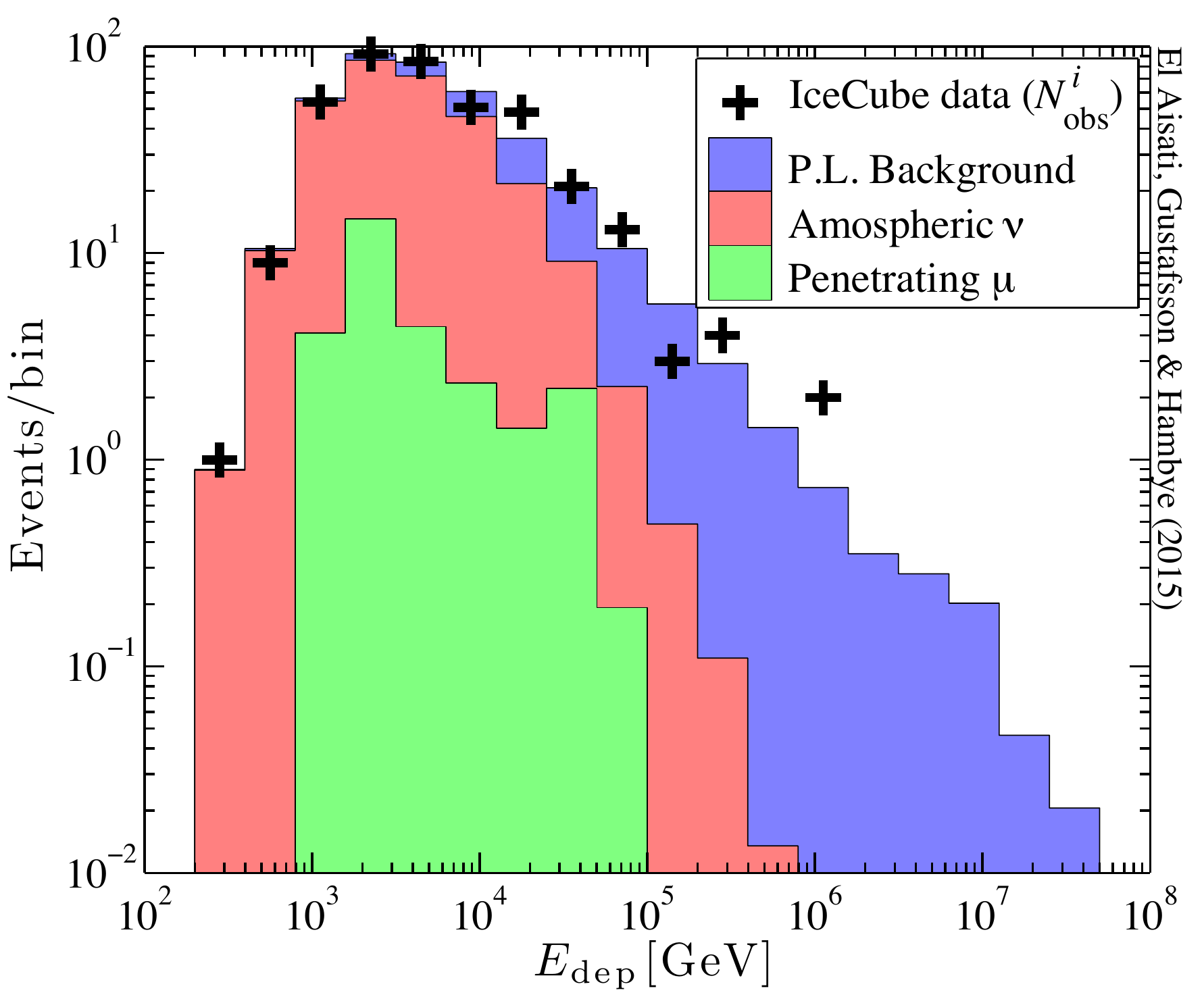}
\caption{Deposited energy spectrum (black $+$) measured by IceCube. The colored regions show the expected atmospheric muon (bottom green) and neutrino (middle red) background components as well as the best-fit astrophysical neutrino power-law background model contribution (top blue).}
\label{fig:ICdata}
\end{figure}

The analysis was performed on a 2-year data sample from the IceCube Collaboration \cite{Aartsen:2014muf}, with energies ranging from $100 $ to $10^8 $ GeV. The distribution of events, as a function of deposited energy $E_\text{dep}$, is shown in figure \ref{fig:ICdata} (black $+$).

In order to see if there is any hint for a smoking-gun in this sample, we use the approach that consists in attempting to reject the background-only hypothesis $\mathcal{H}_0$
using a profile likelihood ratio test statistic \cite{Feldman:1997qc}. 

We model the number of events in each energy bin $i$ with a Poisson probability distribution. Hence, the binned likelihood reads
\be
\mathcal{L} = \prod_{{\rm bins}\; i } \frac{(N^i_\mathrm{model})^{N^i_{\rm obs}}}{N^i_{\rm obs} !} e^{-N^i_\mathrm{model}},
\ee
where $N_\text{obs}^i $ gives the number of observed events in bin $i$, and $N_\text{model}^i $ is the model prediction in that bin. The model includes four contributions in total:
\be
N_\text{model}^i (n_\text{sig}, m_\text{DM}, n_{1,2,3}, \gamma)=  n_\text{sig} N_\text{DM}^i(m_\text{DM})+ n_1 \green{N_\mu^i} + n_2 \red{N_\nu^i} + n_3 \blue{N_\text{astro}^i}(\gamma)
\label{eq:Nmodel}
\ee
Our backgrounds consist of the atmospheric neutrinos (red area in figure \ref{fig:ICdata}), atmospheric muons (green) and neutrinos potentially coming from an astrophysical source that is \textit{not} DM (blue) and whose flux is simply modelled by a power-law with spectral index $-\gamma$. On top of the pre-cited backgrounds, we allow for the injection of a DM signal (first term in Eq.~(\ref{eq:Nmodel})). Taking into account energy dispersion effects, we test for the presence of different types of smoking-guns, that is, different alternative hypotheses $\mathcal{H}_1$: monochromatic neutrinos \cite{Rott:2014kfa}, box-shaped spectra \cite{Guo:2010vy,Garcia-Cely:2016pse} and power-laws with a kinematical cut-off at half of the DM mass $m_\text{DM}$ \cite{Garcia-Cely:2016pse}. A branching ratio of one is always assumed.

We finally build, for each hypothesis $\mathcal{H}_1$, the profile likelihood ratio test statistic

\be
\text{TS}= 2 \ln \frac{ \mathcal{L}(n_{\rm sig} = n_{\rm sig,best}, \hat{\theta} )}{ \mathcal{L}(n_{\rm sig} = 0, \hat{\hat{\theta}})},
 \label{eq:TS} 
\ee
where $\hat{\theta} \equiv (\hat{n}_{1,2,3}, \hat{\gamma})$ maximizes the unconditional $\mathcal{L}$ and $\hat{\hat{\theta}} \equiv (\hat{\hat{n}}_{1,2,3}, \hat{\hat{\gamma}})$ maximizes $\mathcal{L}$ under the condition $n_\text{sig}=0$, that is, the null hypothesis $\mathcal{H}_0$. The tests are carried out at  fixed DM masses $m_\text{DM}$.
The (local) significance for rejecting $\mathcal{H}_0$ in favour of the alternative hypothesis $\mathcal{H}_1$ tested is given  by $\sqrt{\text{TS}}$ (Wilks theorem) \cite{Wilks1938}.

\section{Results}
\label{sec:results}
\begin{figure}[h!]
\centering
\begin{minipage}[c]{.5\linewidth}
\includegraphics[scale=.34]{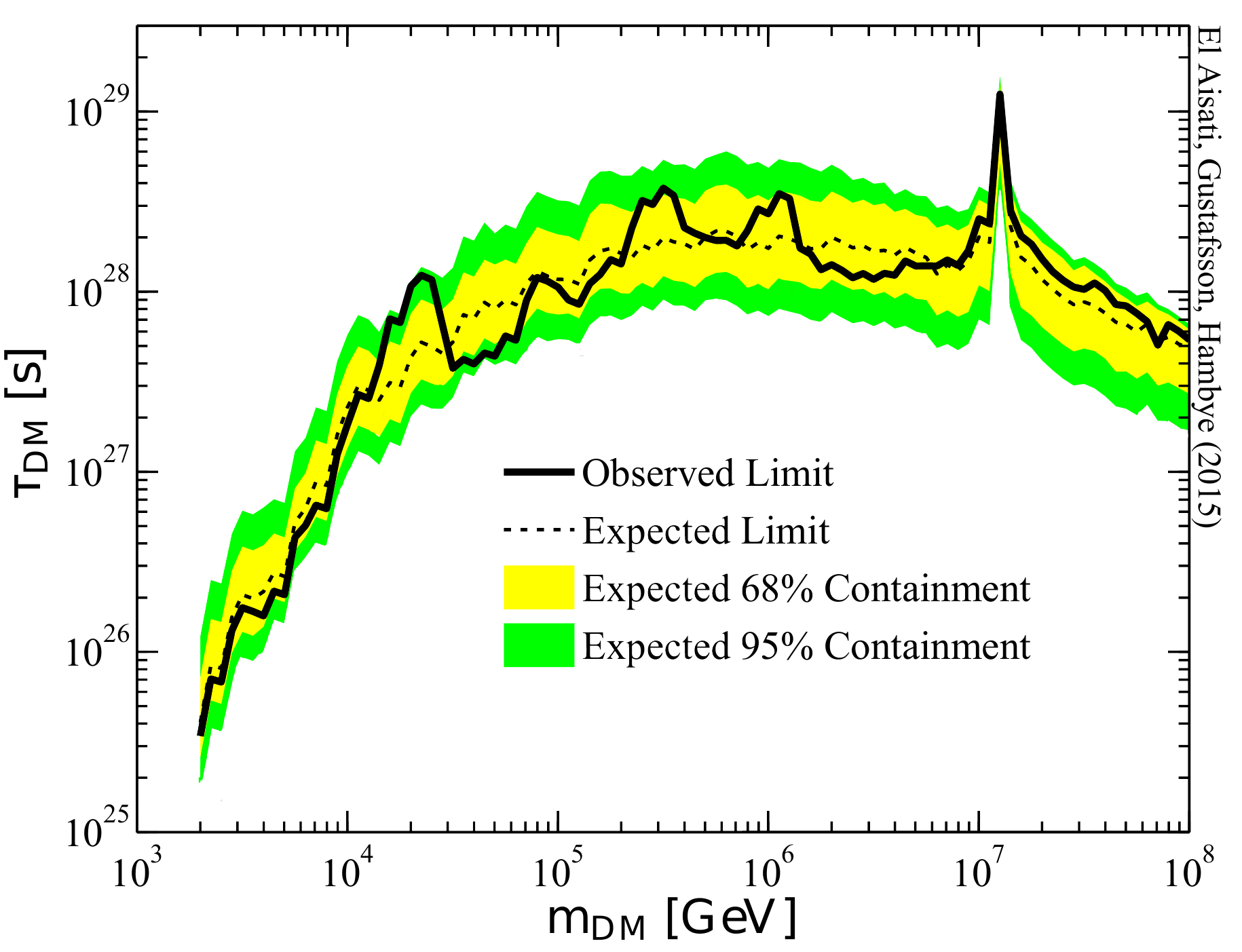}
\caption{95$\%$ C.L. lifetime limits (solid curve) on the DM particle decay lifetime into monochromatic neutrinos. Expected sensitivity reach (dashed curve) and its 68$\%$ (yellow) and 95$\%$ (green) containment bands are also shown.}
\label{fig:lim1}
\end{minipage}
\hfill
\begin{minipage}[c]{.45\linewidth}
\includegraphics[trim={0 9mm 0 0},scale=.37]{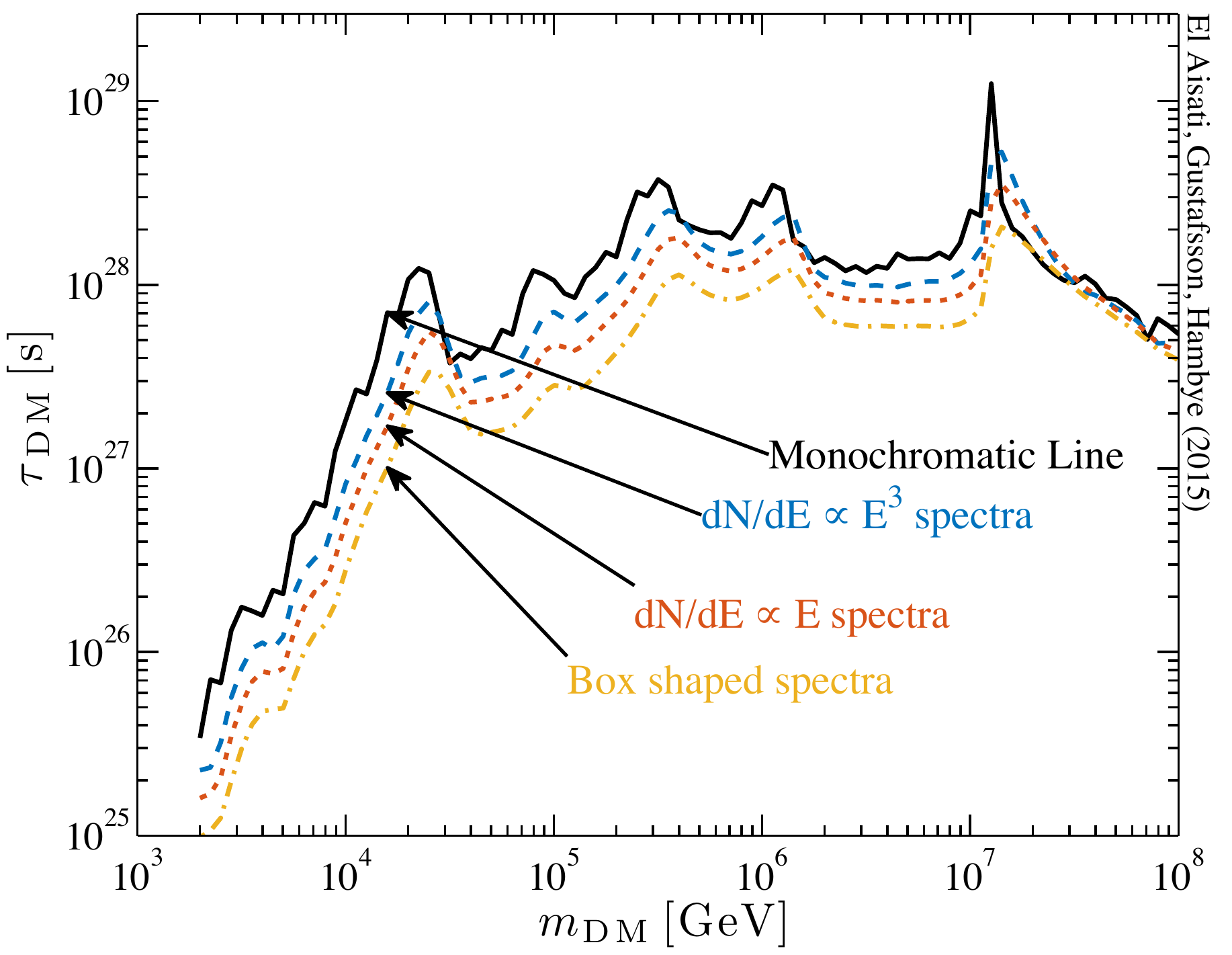}
\caption{95$\%$ C.L. lifetime limits on various sharp spectra from DM decay: monochromatic line, internal bremsstrahlung-like ($dN/dE \propto E, E^3$) and box-like spectrum.}
\label{fig:lim2}
\end{minipage}
\end{figure}

In all of our tests, $\sqrt{\text{TS}}$ is found to be $< 2 \sigma$, locally. The 95\,\% C.L. lower limits on $\tau_{\rm DM}$, as a function of $m_\text{DM},$ are then set by requiring that the profile likelihood $\mathcal{L}(n_{\rm sig})$  with respect to its maximum values to have $\text{TS} < 2.71$. 
They are presented in figure \ref{fig:lim1} in the case of a pure monochromatic neutrino line at the decay (together with the $68 \%$ and $95 \%$ containment bands) and in figure \ref{fig:lim2} for few other spectra assumptions. Note the effect of the Glashow resonance on the limits around $m_\text{DM} = 2 \times 6.3 ~\text{PeV}$.

\begin{wrapfigure}[15]{r}{8cm}
\vspace{-1.1cm}
\includegraphics[scale=.57]{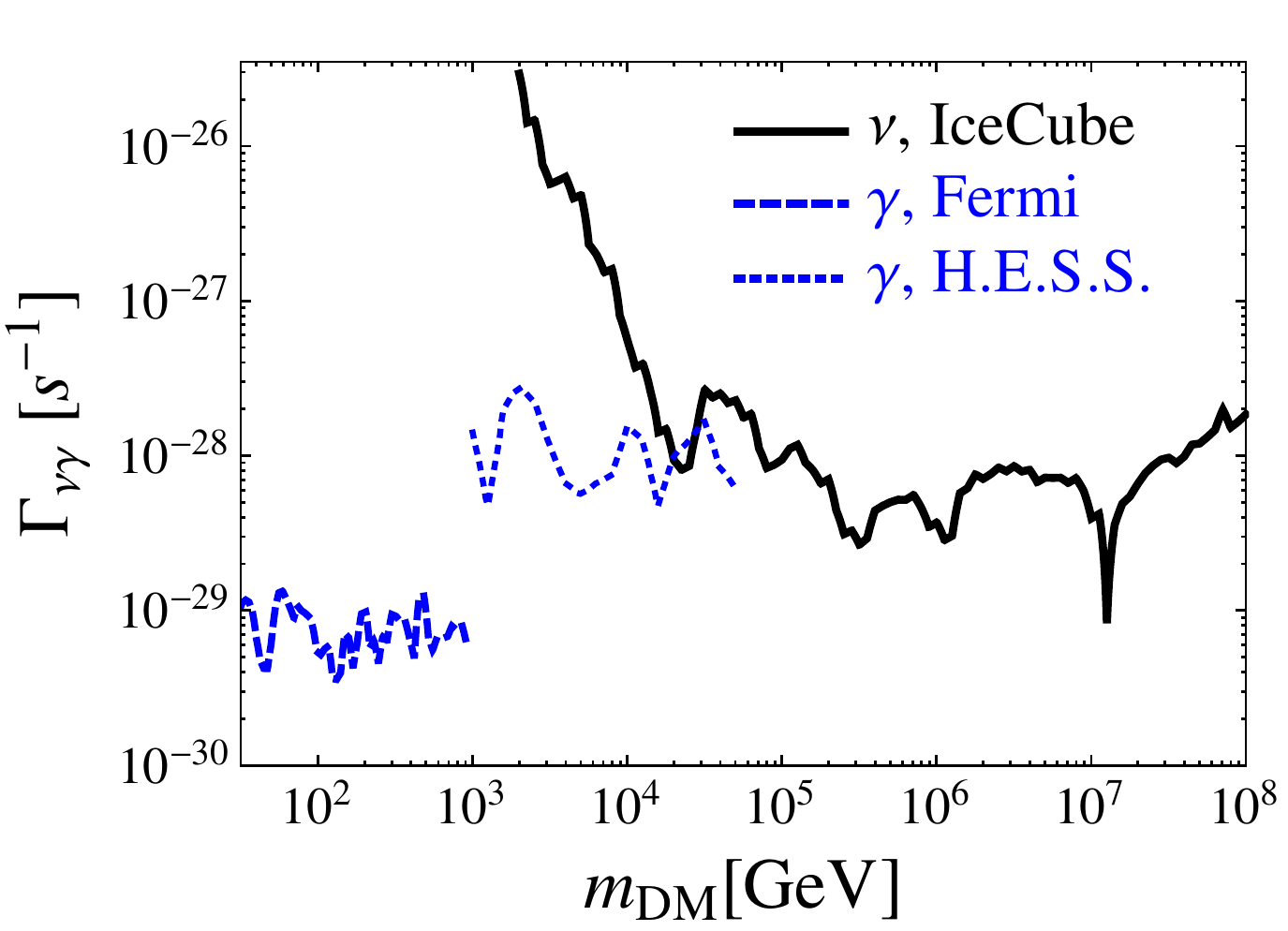}
\caption{Limits on DM decay width $\Gamma(DM\rightarrow \gamma X)$ from Fermi-LAT (dashed blue line) and H.E.S.S. (dotted blue), and on $\Gamma(DM\rightarrow \nu X)$ (solid black), assuming $X$ is not producing a signal in any of the detectors.}\label{fig:ICfermiHESS}
\end{wrapfigure} 

\vspace{.2cm}
It is interesting to compare the limits obtained from neutrino-line searches to the ones from gamma-ray line searches (Fermi-LAT \cite{Ackermann:2015lka} and H.E.S.S. \cite{Abramowski:2013ax,Gustafsson:2013gca}), as is done in figure \ref{fig:ICfermiHESS}. 
For multi-TeV energies, we show that the sensitivities reached on neutrino lines are now comparable to those existing on monochromatic gamma-ray lines. For example, around 20 TeV DM masses, we have improved them by about a factor 50 with respect to the latest IceCube limits available for DM decay \cite{Abbasi:2011eq}. 

Note also the strong constraints obtained from the neutrino sky above the maximum energy considered by H.E.S.S ($\sim$25 TeV), and where---to our knowledge---there are no numerically relevant gamma-line constraints.

\vspace{.8cm}
With foreseen improvements in both neutrino  \cite{Aartsen:2013mla,Aartsen:2014njl,Avrorin:2013uyc,Aartsen:2014oha,Margiotta:2014eaa} and gamma-ray \cite{Bringmann:2012ez,Bergstrom:2012vd,Conrad:2015bsa,Ibarra:2015tya} data, this opens up increased chances to see a ``double smoking gun'' signal in the form of a monochromatic neutrino line plus a gamma-ray line from DM particles' decays \cite{Aisati:2015ova}.

\vspace{.5cm}
 
The same analysis can of course be performed under the assumption of annihilating DM, with the slight difference that limits would be set on the average annihilation cross section of DM particles $\langle \sigma_\text{ann} v \rangle$. We do not show the corresponding limits here, as we are currently investigating this option in a more detailed study, including angular information of events in our likelihoods.

\section{Summary}
A search for sharp neutrino features from DM decay, using a binned likelihood in energy, has been presented. No signal was found, but we have illustrated with a particular example---DM decay---the potential of neutrinos feature searches in the context of DM: that is, the possibility to achieve, in regions of the parameter space, sensitivities comparable with gamma-ray counterparts. 
In particular, the discovery of a ``double-barrelled'' smoking gun in the future does not seem unrealistic at all.

\section*{Acknowledgments}
I would like to thank the organizers of the 51st Rencontres de Moriond for giving me the opportunity to present this work, as well as for financial support.
I also thank M.~Gustafsson and T.~Hambye for their collaboration. This work is supported by the FRIA, Invisibles, the FNRS-FRS, the IISN and the Belgian Science Policy, IAP VI-11.

\end{document}